\documentclass[12pt]{article}

\usepackage{amsfonts}
\usepackage{amsmath}
\usepackage{epsfig}
\usepackage{latexsym}
\usepackage{graphicx}
\usepackage{color}
\usepackage{amssymb}
\usepackage{epstopdf}
\numberwithin{equation}{section}
\allowdisplaybreaks

\usepackage{url}

\def\be{\begin{equation}}
\def\ee{\end{equation}}
\def\bea{\begin{eqnarray}}
\def\eea{\end{eqnarray}}
\def\bequ{\begin{equation}}
\def\eequ{\end{equation}}

\def\del{\partial}

\def\delS{\partial_\#^d}
\def\delSS {  \partial_\#^{2d}  }

\renewcommand{\thefootnote}{\fnsymbol{footnote}}

\newcommand{\eq} {equation}
\newcommand{\eqa} {eqnarray}
\newcommand{\NN} {\nonumber}

\begin{document}
\hfuzz=100pt
\title{Scalar, fermionic and supersymmetric field theories\\ 
with subsystem symmetries in $d+1$ dimensions}

\author{
\large 
Masazumi Honda$^{1,2}$\footnote{masazumi.honda(at)yukawa.kyoto-u.ac.jp}, \quad
Taiichi Nakanishi$^{1}$\footnote{taiichi.nakanishi(at)yukawa.kyoto-u.ac.jp}
\vspace{1em} \\
\\
$^{1}${\small\it Center for Gravitational Physics and Quantum Information,} \\
{\small\it Yukawa Institute for Theoretical Physics, Kyoto University,}\\
{\small\it  Sakyo-ku, Kyoto 606-8502, Japan}
\\
$^{2}${\small\it Interdisciplinary Theoretical and Mathematical Sciences Program (iTHEMS), }\\
{\small\it RIKEN, Wako 351-0198, Japan}
}

\date{\small{December 2022}}

\maketitle
\thispagestyle{empty}
\centerline{}

\begin{abstract}
We study various non-relativistic field theories with exotic symmetries called subsystem symmetries,
which have recently attracted much attention in the context of fractons. 
We start with a scalar theory called $\phi$-theory in $d+1$ dimensions
and discuss its properties studied in literature for $d\leq 3$
such as self-duality, vacuum structure, 't Hooft anomaly, anomaly inflow and lattice regularization.
Next we study a theory called chiral $\phi$-theory
which is an analogue of a chiral boson with subsystem symmetries.
Then we discuss theories including fermions with subsystem symmetries.
We first construct a supersymmetric version of the $\phi$-theory and
dropping its bosonic part 
leads us to
a purely fermionic theory with subsystem symmetries called $\psi$-theory.
We argue that lattice regularization of the $\psi$-theory generically suffers 
from an analogue of doubling problem as previously pointed out in the $d=3$ case.
We propose an analogue of Wilson fermion to avoid the ``doubling" problem.
We also supersymmetrize the chiral $\phi$-theory and
dropping the bosonic part again gives us a purely fermionic theory.
We finally discuss vacuum structures of the theories with fermions and 
find that they are infinitely degenerate because of spontaneous breaking of subsystem symmetries.

\end{abstract}

\vfill
\noindent

{\small YITP-22-135, RIKEN-iTHEMS-Report-22}

\renewcommand{\thefootnote}{\arabic{footnote}}
\setcounter{footnote}{0}

\newpage
\pagenumbering{arabic}
\setcounter{page}{1}
\tableofcontents

\section{Introduction}
Fracton phases have recently attracted much attention in various contexts.
While previous attention was mainly from viewpoints of quantum information and condensed matter physics\footnote{
See \cite{Nandkishore:2018sel,Pretko:2020cko} for reviews.
},
there have recently appeared many works related to high energy physics.
Fracton phases are featured by a new type of symmetry called subsystem symmetry \cite{Vijay:2015mka,Paramekanti2002,Chamon:2004lew,Haah:2011drr}.
which
is a generalization of global symmetry
in the sense that symmetry operators partially violate topological property.
In this paper 
we generalize various continuum field theories with subsystem symmetries in literature
and study their properties,
mainly based on the framework developed in \cite{Seiberg:2019vrp,Seiberg:2020bhn,Seiberg:2020wsg,Seiberg:2020cxy,Gorantla:2020xap,Gorantla:2020jpy,Rudelius:2020kta,Gorantla:2021svj,Gorantla:2021bda,Yamaguchi:2021qrx,Burnell:2021reh,Yamaguchi:2021xeq}.
In particular
we discuss various basic properties 
such as symmetries, self-duality, vacuum structure, 't Hooft anomaly, anomaly inflow and lattice regularization.
Our class of theories include ones with fermions as well as supersymmetric theories
and
we keep spatial dimensions as general as possible
while most of studies on fractons have focused 
only on bosonic theories in $2+1$ and $3+1$ dimensions so far\footnote{
Some exceptions are \cite{You:2018bmf,Tantivasadakarn:2020lhq,Shirley:2020ass,Yamaguchi:2021qrx,Razamat:2021jkx,Geng:2021cmq,Distler:2021qzc,Katsura:2022xkg,Cao:2022lig}.
}.
Generalization of fracton-like theories would lead us to
deeper understanding, new connections to other physics and
some insights on experimental realizations of fractons.

All the theories studied in this paper are non-relativistic theories
except for $1+1$ dimensional cases.
In particular, a space derivative in their Lagrangians in $d+1$ dimensions appears only in the combination
\begin{\eq}
\del_1 \del_2 \cdots \del_d ,
\label{eq:derivative}
\end{\eq}
while time derivative appears in a standard way.
Therefore a spatial rotation symmetry of the theories is not continuous and
consists only of discrete 90 degree rotations. 
Because of the exotic derivative \eqref{eq:derivative},
a dispersion relation for a classical particle in the class of theories also become exotic:
\begin{\eq}
E^2 \propto \left( p_1 p_2 \cdots p_d \right)^2 ,
\label{eq:dispersion}
\end{\eq}
which tells us that the energy $E$ can be small 
when only one component of the momentum $(p_1 ,\cdots ,p_d )$ is sufficiently small
even if other components are large.
This feature makes an analysis of vacuum structures complicated even in free theories.

We start with a compact scalar theory called $\phi$-theory in $d+1$ dimensions,
where the $d=1$ case corresponds to the standard compact boson.
This theory for $d=2,3$ has been already studied well \cite{Seiberg:2020bhn,Gorantla:2020xap}
while generic $d$ case has not been explicitly studied in literature.
We discuss various properties of the $\phi$ theory
such as symmetries, self-duality, vacuum structure, 't Hooft anomaly, anomaly inflow and lattice regularization.
Next we study a theory called chiral $\phi$-theory
which is a generalization of the chiral boson to $d+1$ dimensions with $d$ odd.
We will see that 
the chiral $\phi$-theory is related to the $\phi$-theory
in a similar way to the relation between the standard chiral boson and compact boson
as in the $d=3$ case \cite{Burnell:2021reh,Yamaguchi:2021xeq}.
We also discuss vacuum structures and 't Hooft anomaly of the chiral $\phi$-theory.

Then we turn to theories with fermions.
Following the strategy of \cite{Yamaguchi:2021qrx} for $d=3$, 
we first construct a supersymmetric (SUSY) version of the $\phi$-theory for general odd $d$.
This is a generalization of 
the supersymmetrization of the standard compact boson with $\mathcal{N}=(1,1)$ SUSY 
in $1+1$ dimensions.
We discuss basic properties of the theory
such as subsystem symmetries, self-duality and coupling to gauge field
in terms of a superfield formalism.
Dropping the bosonic part of the supersymmetric $\phi$-theory
enables us to find a purely fermionic theory with subsystem symmetries. 
This theory is a natural generalization of the $\psi$-theory 
introduced for $d=3$ in \cite{Yamaguchi:2021qrx} to general odd $d$.
We also argue that a lattice regularization of the $\psi$-theory generically suffers from an analogue of doubling problem as pointed out in the $d=3$ case \cite{Yamaguchi:2021qrx}.
We will see that 
the number of ``doublers" in a naive lattice fermion theory is 
much more complicated than usual lattice fermion thories.
To avoid the ``doubling" problem,
we propose an analogue of Wilson fermion which makes the ``doubler modes" infinitely massive in the continuum limit.

It turns out that there is a similar story for the chiral $\phi$-theory
which has not been discussed in literature.
We supersymmetrize the chiral $\phi$-theory 
in a similar way to the supersymmetization of the chiral boson with $\mathcal{N}=(1,0)$ SUSY in $1+1$ dimensions \cite{Bellucci:1987mj,Sonnenschein:1988ug}.
Then dropping its bosonic part again gives rise to a purely fermionic theory
which is an analogue of the standard chiral fermion in $1+1$ dimensions.
Finally we discuss vacuum structures of the above theories including fermions.
We find that
the theories with periodic boundary condition have infinitely many degenerate vacua 
except $d=1$ as in the $d=3$ case \cite{Yamaguchi:2021qrx}.
The degeneracy essentially comes from infinitely many zero modes of the fermions
associated with spontaneous symmetry breaking of subsystem symmetries.

This paper is organized as follows.
In sec.~\ref{sec:phi-theory},
we discuss scalar theories with subsystem symmetries in $(d+1)$ dimensions.
Most of sec.~\ref{sec:phi-theory}
is essentially a straightforward generalization of \cite{Seiberg:2020bhn,Gorantla:2020xap,Burnell:2021reh} 
which discussed properties of the scalar theories with subsystem symmetries for the $d=2$ and $d=3$ cases.
In sec.~\ref{sec:SUSY},
we construct a supersymmetric version of the $\phi$-theory and
extract a purely fermionic theory with subsystem symmetries.
We also discuss a lattice regularization of the fermionic theory.
In sec.~\ref{sec:chiralSUSY},
we construct a supersymmetric version of the chiral $\phi$-theory.
In sec.~\ref{sec:vacuum},
we discuss vacuum structures of the theories with fermions.
Sec.~\ref{sec:conclusion} is devoted to conclusion and discussions.

\section{Scalar theories with subsystem symmetries}
\label{sec:phi-theory}
In this section
we discuss 
two scalar theories with subsystem symmetries 
called the $\phi$-theory and chiral $\phi$-theory in $d+1$ dimensions.
Sec.~\ref{sec:phi} for the $\phi$-theory is essentially
a straightforward generalization of \cite{Seiberg:2020bhn}
and \cite{Gorantla:2020xap},
where $d$ was fixed to $d=2$ and $d=3$ respectively.
Sec.~\ref{sec:chiral-phi} for the chiral $\phi$-theory 
is basically an extension of \cite{Burnell:2021reh},
where $d$ was fixed to $d=3$.

\subsection{$\phi$-theory}
\label{sec:phi}
The $\phi$-theory is a non-relativistic theory of a real compact scalar $\phi$ 
with a periodicity described below.
We take a spacetime coordinate as $(t, x^1 ,\cdots ,x^d )$
where $t$ is a usual time direction and $x^i$ ($i =1,\cdots ,d$) is an ``exotic" spatial direction.
The Lagrangian of the $\phi$-theory is 
\begin{\eqa}
\mathcal{L}_\phi [\phi ]
= \frac{\mu_0}{2} \Bigl[ (\del_t \phi )^2 -\alpha^2 ( \delS \phi )^2 \Bigr] ,
\label{eq:original}
\end{\eqa} 
where $\mu_0$ and $\alpha$ are constants, 
and 
$\del_\#$ is a spatial differential operator defined by
\begin{\eq}
\del_\#^d := \del_1 \cdots \del_d .
\end{\eq}
The scalar field $\phi$ is subject to the following identification
\begin{\eq}
\phi (t,x) \sim \phi (t,x) +2\pi \sum_{i=1}^d w^i (x;\hat{x}^i ) ,
\label{eq:identify}
\end{\eq}
where $w^i \in \mathbb{Z}$ is any integer-valued discontinuous function and
the symbol $(x; \hat{x}^i )$ denotes 
dependence on the spatial coordinates $(x^1 ,\cdots ,x^d )$ except $x^i$.

The $\phi$-theory for $d=1$ is nothing but the standard compact boson theory in $1+1$ dimensions
that is of course relativistic while higher $d$ case is non-relativistic.
There is a technical but important difference between odd and even $d$
that is a ``Leibniz rule" for the differential operator $\delS$: 
\begin{\eq}
\left( \delS \phi_1 (x) \right) \phi_2 (x) 
\simeq  (-1)^d \phi_1 (x) \left( \delS \phi_2 (x) \right) ,
\label{eq:Leibniz}
\end{\eq}
where $\simeq$ denotes equality up to total derivative terms.
Noting this property,
the equation of motion for $\phi$ is given by
\begin{\eq}
\del_t^2 \phi +(-1)^d \alpha^2 \delSS  \phi =0  ,
\label{eq:EOM-phi}
\end{\eq}
where $\delSS := (\delS)^2$.
It is also convenient to use the following differential operator 
\begin{\eq}
\del_\pm  := \frac{1}{2} \left( \del_t \pm \alpha \delS  \right) ,
\end{\eq}
In terms of $\del_\pm$, we can also rewrite the Lagrangian as
\begin{\eq}
\mathcal{L}_\phi [\phi ]
= 2\mu_0 ( \del_- \phi ) (\del_+ \phi ) .
\end{\eq}
In particular, 
the equation of motion for odd $d$ is written as\footnote{
For even $d$, it is $(\del_+^2 +\del_-^2 )\phi =0$.
}
\begin{\eq}
\del_+ \del_- \phi =0  \quad {\rm for\ odd\ } d .
\end{\eq}

\subsubsection{Subsystem symmetries}
The $\phi$-theory is known to have two types of subsystem symmetries.
One is called ``momentum multi-pole symmetry''.
This essentially comes from a reinterpretation of the e.o.m.~\eqref{eq:EOM-phi}
as a conservation law:
\begin{\eq}
\del_t P_t -\delS P_\# =0 ,
\label{eq:momentum_conserve}
\end{\eq}
where
\begin{\eq}
P_t := \mu_0 \del_t \phi ,\quad 
P_\# =   (-1)^{d+1} \mu_0 \alpha^2  \delS \phi .
\end{\eq}
Associated with this, we have the following conserved charges
\begin{\eq}
Q_m^i (x; \hat{x}^i )
:= \int dx^i P_t ,
\end{\eq}
which generates a shift of $\phi$ by any real-valued function independent of $t$ and $x^i$:
\begin{\eq}
Q_m^i (x;\hat{x}^i ) : \phi (t,x)\ \rightarrow\ \phi (t,x) +f^i (x;\hat{x}^i ).
\label{eq:trans_boson}
\end{\eq}
Because of the identification \eqref{eq:identify}, this symmetry is $U(1)$ rather than $\mathbb{R}$ 
and \eqref{eq:identify} can be viewed as a ``large" transformation.
Note also that not all of $Q_m^i $'s are independent of each other and we have some overlaps:
\begin{\eq}
\int dx^j Q_m^i (x; \hat{x}^i ) = \int dx^i Q_m^j (x; \hat{x}^j ) ,
\label{eq:Gauss}
\end{\eq}
which generate a shift of $\phi$ independent of $t$, $x^i$ and $x^j$.

The other symmetry is called ``winding multi-pole symmetry'' and
its conservation law comes from the trivial identity
\begin{\eq}
\del_t W_t^\# - \delS  W =0 ,
\end{\eq}
where
\begin{\eq}
W_t^\# = \frac{1}{2\pi} \delS \phi ,\quad W= \frac{1}{2\pi} \del_t \phi  .
\end{\eq}
Associated with this, we have the following conserved charges
\begin{\eq}
Q_w^i (x; \hat{x}^i )
:= \int dx^i W_t^\#  .
\end{\eq}
In contrast to the momentum symmetry,
this symmetry is independent of details of Lagrangian and an analogue of ``topological symmetry"
coming from a generalization of ``Bianchi identity" for $\phi$.

\subsubsection{Self-duality}
\label{sec:duality_phi}
The $\phi$-theory is also known to have a self-duality as in the standard compact boson.
Here we derive the self-duality for any $d$ 
in a slightly different way from the argument for $d=2$ and $d=3$ \cite{Seiberg:2020bhn,Gorantla:2020xap}.
Let us regard $(\del_t \phi ,\delS \phi )$ as dynamical variables rather than $\phi$ itself and
consider the Lagrangian:
\begin{\eq}
\mathcal{L}_1 \left[ \del_t \phi ,\delS \phi  , \tilde{\phi} \right]
=\frac{\mu_0}{2} \Bigl[ (\del_t \phi )^2  -\alpha^2 (\delS \phi )^2 \Bigr]
- \tilde{\phi}  \left( \del_t W_t^\#  -  \delS  W  \right)  ,
\end{\eq}
where $\tilde{\phi}$ is a Lagrange multiplier 
to impose the conservation law of the winding symmetry.
One can go back to the original Lagrangian \eqref{eq:original}
just by integrating $\tilde{\phi}$ out.
Integrating by parts and completing square, we find
\begin{\eqa}
\mathcal{L}_1
&\simeq &\frac{\mu_0}{2} \Bigl[ (\del_t \phi )^2  -\alpha^2 (\delS  \phi )^2 \Bigr] 
+\frac{1}{2\pi} (\del_t \tilde{\phi} ) (\delS \phi)
+\frac{1}{2\pi} (-1)^d (\delS \tilde{\phi} )  (\del_t \phi ) \NN\\
&= & \frac{\mu_0}{2} \left(\del_t \phi +\frac{1}{2\pi\mu_0}(-1)^d  \delS \tilde{\phi} \right)^2
-\frac{\mu_0\alpha^2}{2} \left( \delS \phi -\frac{1}{2\pi\mu_0 \alpha^2}  \del_t \tilde{\phi} \right)^2 \NN\\
&& +\frac{1}{8\pi^2 \mu_0 \alpha^2 }\Bigl[
 (\del_t \tilde{\phi} )^2 -\alpha^2 ( \delS \tilde{\phi} )^2 \Bigr] . 
\end{\eqa}
Then, integrating out $\del_t \phi$ and $\delS \phi$ leads us to
\begin{\eqa}
\mathcal{L}_2 [\tilde{\phi} ]
= \frac{\tilde{\mu}_0 } {2 }
\Bigl[
 (\del_t \tilde{\phi} )^2 -\alpha^2 ( \del_1 \cdots \del_d \tilde{\phi} )^2 \Bigr] ,
\end{\eqa}
where
\begin{\eq}
\tilde{\mu}_0 := \frac{1}{(2\pi \alpha )^2 \mu_0} .
\end{\eq}
This takes the same form as the original Lagrangian \eqref{eq:original}
but with a different value of the overall coefficient in the Lagrangian,
which is physically a radius of the target space $S^1$.
Therefore the $\phi$-theory with the parameters $(\mu_0 ,\alpha )$ 
is self dual to the one with $(\tilde{\mu}_0 ,\alpha )$.
From the above derivation, 
we have the following duality relations for the fields:
\begin{\eq}
\del_t \phi  =\frac{1}{2\pi\mu_0} (-1)^{d+1}  \delS \tilde{\phi} ,\quad
\delS \phi = \frac{1}{2\pi\mu_0 \alpha^2}  \del_t \tilde{\phi} .
\label{eq:tilde}
\end{\eq}
This can be also written as
\begin{\eq}
P_t   =  (-1)^{d+1}\tilde{W}_t^\#  ,\quad P_\#  =  \frac{(-1)^{d+1}}{\alpha^2} \tilde{W} ,
\end{\eq}
and
\begin{\eq}
W = \alpha^2 \tilde{P}_\#  ,\quad W_t^\#  = \tilde{P}_t .
\end{\eq}
Thus, 
the duality is an exchange of momentum and winding
as in the familiar $T$-duality for the compact scalar in $1+1$ dimensions.

\subsubsection{Vacuum structure on torus}

Let us take the space to be 
a $d$-dimensional torus $S_{\ell_1}^1 \times \cdots \times S_{\ell_d}^1$.
The dispersion relation for a classical particle of the model is $E \propto p_1 \cdots p_d$ and
this implies that even if some of $p_j$'s are large,
another small $p_j$ can make the energy a finite value including zero.
While this fact seems to suggest spontaneous breaking of the momentum subsystem symmetries,
it is known that for $d=2$ and $d=3$,
this is a classical feature and
the symmetries are unbroken quantum theoretically \cite{Seiberg:2020bhn,Gorantla:2020xap}.
One can show that this is true for general $d\geq 2$ as follows.

To see a vacuum structure of the theory, we first switch to Hamilton formalism:
\begin{\eq}
H =\int d^d x \Biggl[ \frac{1}{2\mu_0} \Pi^2 +\frac{\mu_0 \alpha^2}{2} (\delS \phi )^2 \Biggr] ,
\label{eq:Hamiltonian_phi}
\end{\eq}
where $\Pi$ is the conjugate momentum of $\phi$ given by
\begin{\eq}
\Pi := \mu_0 \del_t \phi ,
\end{\eq}
and the operators satisfy the canonical commutation relation
\begin{\eq}
 [\phi (x) , \Pi (y) ] = i \delta^{(d)} (x-y) .
\end{\eq}
If we make the Fourier expansions
\begin{\eq}
\phi (x) = \sum_{ \{ p_j \} } \phi_{ \{ p_j\} } e^{ i p_j x_j   } ,\quad
\Pi (x) = \sum_{ \{ p_j \} } \Pi_{ \{ p_j\} } e^{ i p_j x_j   } ,
\end{\eq}
then we find
\begin{\eq}
H =   \sum_{ \{ p_j \} } \Biggl[ \frac{1}{2\mu_0} | \Pi_{ \{p_j \} } |^2  
 +\frac{\mu_0 \alpha^2 (p_1 \cdots p_d )^2 }{2} | \phi_{ \{p_j \} } |^2 \Biggr] .
\end{\eq}
Note that the second term vanishes when either one of $\{ p_j \}$ is zero.
These modes are called momentum modes and we will treat them separately.
For the modes with all nonzero momenta,
we can simply rewrite them in terms of creation/annihilation operators and
therefore 
the ground states are necessarily the Fock vacua for these modes. 
Thus we can focus on the momentum modes.
Going back to the coordinate representation,
the momentum mode part of $\Pi (x)$ is given by
\begin{\eq}
\sum_{j=1}^d  \frac{1}{\ell_j}\oint dx^j \Pi (x)
= \sum_{j=1}^d  \frac{\mu_0}{\ell_j} Q_m^j  (x;\hat{x}^j ) .
\end{\eq}
Then the relevant part of the Hamiltonian is
\begin{\eq}
\left. H \right|_{\rm momentum\ modes} 
= \int d^d x  \Biggl[  \sum_{j=1}^d  \frac{1}{\ell_j} Q_m^j  (x;\hat{x}^j )  \Biggr]^2 ,
\end{\eq}
whose eigenstates are ones of the multi-pole momentum charge $Q_m^i  (x;\hat{x}^i )$ for all $i$.

We also have to take the identification \eqref{eq:identify} into account.
This is done by imposing that physical states are invariant under the large gauge transformation \eqref{eq:identify}:
\begin{\eq}
 e^{2\pi i \int d^{d-1}\hat{x}_j  w_j (x; \hat{x}^j ) Q_m^j (x:\hat{x}^j ) } |{\rm phys}\rangle =|{\rm phys}\rangle ,
\end{\eq}
where $d^{d-1}\hat{x}_j := \prod_{i=1 , i\neq j}^d dx_i $. 
This is parallel to the free one-particle quantum mechanics on $S^1$.
The above condition is solved by
\begin{\eq}
Q_m^j (x:\hat{x}^j ) |{\rm phys}\rangle 
= \sum_\alpha N_\alpha^j \ \delta^{(d-1)} (x- x_\alpha ; \hat{x}^j )   |{\rm phys}\rangle ,
\end{\eq}
with $N_\alpha^j \in \mathbb{Z}$.
Furthermore the overlap condition \eqref{eq:Gauss}, which can be viewed as a Gauss law, requires
\begin{\eq}
\sum_\alpha N_\alpha^j  =\sum_\beta N_\beta^k \quad {\rm for}\ ^\forall j,k .  
\end{\eq}
The above conditions imply that
the energy of any physical state with nonzero multi-pole momentum charges is divergent
as we encounter a bunch of $\delta (0)$'s.
Thus, the ground state of the $\phi$-theory on torus has vanishing multi-pole momentum charges and unique.
In the infinite volume limit,
the modes with non-zero $p_1 \cdots p_d$ can have energies smoothly connected to the ground state energy and
therefore the theory is gapless in this sense.

\subsubsection{'t Hooft anomaly and anomaly inflow}
It was argued in \cite{Burnell:2021reh} that
the $\phi$-theory in $2+1$ dimensions has
a mixed 't Hooft anomaly between the $U(1)$ multi-pole momentum and winding symmetries
as in the 
usual compact boson in $1+1$ dimensions.
Here we argue that
the $\phi$-theory for general $d$ has essentially the same 't Hooft anomaly
by simply extending the argument of \cite{Burnell:2021reh}.

To study the 't Hooft anomaly, we shall first couple the $\phi$-theory 
to background gauge fields of the symmetries.
Let us work in Euclid signature and 
denote the background gauge fields for the momentum and winding symmetries 
by $(A_\tau ,A_\# )$ and $(\tilde{A}_\tau ,\tilde{A}_\# )$ respectively.
The Lagrangian after the coupling is\footnote{
One might wonder why the couplings between the gauge fields and currents take 
the form of $A_\tau J_\tau +(-1)^{d+1}A_\# J_\#$.
This is required by imposing that gauge transformation of this term is proportional 
to the current conservation law.
See e.g.~sec.2.4 of \cite{Seiberg:2020wsg}.
}
\begin{\eq}
\mathcal{L}_\phi [ A ,\tilde{A} ]
= \frac{\mu_0}{2} (\del_\tau \phi -A_\tau )^2 +\frac{1}{2\mu} ( \del_\#^d \phi -A_\# )^2  
+\frac{i}{2\pi} \Bigl[ \tilde{A}_\tau (\delS \phi -A_\# ) 
+(-1)^{d+1} \tilde{A}_\# (\del_\tau \phi - A_\tau ) \Bigr] .
\end{\eq} 
We easily see that the action is invariant under the background gauge transformation for the momentum symmetry:
\begin{\eq}
\phi \ \rightarrow \ \phi  +f (\tau ,x) ,\quad
A_\tau \ \rightarrow \ A_\tau +\del_\tau f (\tau ,x),\quad A_\# \ \rightarrow \ A_\# + \delS f (\tau ,x) .
\end{\eq}
However,
it is not invariant under the transformation for the winding symmetry:
\begin{\eq}
\tilde{A}_\tau \ \rightarrow \ \tilde{A}_\tau +\del_\tau \tilde{f} (\tau ,x),\quad 
\tilde{A}_\# \ \rightarrow \ \tilde{A}_\# + \delS \tilde{f} (\tau ,x) ,
\end{\eq}
and the action changes as
\begin{\eq}
S_\phi [ A ,\tilde{A} ] \ \rightarrow\ 
S_\phi [ A ,\tilde{A} ]
+\frac{i}{2\pi} \int d\tau d^d x\ \tilde{f}(\tau ,x) \left(  \del_\tau A_\#  - \delS A_\tau \right) .
\end{\eq} 
This change cannot be removed by adding local counter terms
while keeping the (gauged) momentum symmetry.
This signals a mixed 't Hooft anomaly between the $U(1)$ multi-pole momentum and winding symmetries.
Therefore the vacuum structure of the $\phi$-theory cannot be trivial and
this is consistent with the fact that the $\phi$-theory is a gapless theory.

As in the $d=1$ and $d=2$ cases,
we can find an anomaly inflow argument for the above 't Hooft anomaly.
Let us consider a $(d+2)$-dimensional gauge theory on $S^1 \times \mathbb{R}_{z\geq 0} \times \Sigma_d$
with the Lagrangian
\begin{\eq}
\mathcal{L}_{\rm CS} [A,\tilde{A} ]
= -\frac{i}{2\pi} \Biggl[
 \tilde{A}_\tau ( \del_z A_\# -\delS A_z )
+\tilde{A}_z (\delS A_\tau -\del_\tau A_\# )
+(-1)^d \tilde{A}_\# (\del_z A_\tau -\del_\tau A_z ) 
\Biggr] ,
\end{\eq}
which is similar to a $U(1)$ Chern-Simons theory in three dimensions.
This theory has been studied for $d=3$ in \cite{Burnell:2021reh,Yamaguchi:2021xeq,Yamaguchi:2022apr}.
One can show that
gauge transformation of the Lagrangian is total derivative and
therefore the action is gauge invariant for space without boundary.
For a space with a boundary at $z=0$, the action changes as  
\begin{\eqa}
\delta S_{\rm CS} [A,\tilde{A} ]
&=& -\frac{i}{2\pi} \int dt dz d^d x\  \del_z \left(  \tilde{f} (\delS A_\tau -\del_\tau A_\# ) \right)
\NN\\
&=& -\frac{i}{2\pi} \int dt d^d x \Bigl[ \tilde{f} (\del_\tau A_\# -\delS A_\tau )  \Bigr]_{z=0} ,
\end{\eqa}
which cancels the 't Hooft anomaly.

\subsubsection{Comments on interactions}
In general we can also add interactions
but it is highly constrained if we impose the subsystem momentum symmetries.
For instance, any potential of $\phi$ as in a usual relativistic scalar theory is not invariant under a nontrivial shift of $\phi$.
While it is not easy to classify interactions fully preserving the momentum symmetry,
a simple way is to add a function of $\delS \phi$ :
\begin{\eq}
W\left( \del_\#^d \phi \right) .
\end{\eq}
For this case, the equation of motion for $\phi$ is modified as
\begin{\eq}
\del_t^2 \phi 
-\delS \Bigl[
(-1)^{d+1} \alpha^2 \delS  \phi  +(-1)^d  W' (\delS \phi) \Bigr] =0  .
\end{\eq}
This takes the form of the conservation law \eqref{eq:momentum_conserve}
of the momentum symmetry with a modification of $P_\#$
while $P_t$ remains as the same expression.

If we allow a partial breaking of the symmetry,
there are more options.
For instance,
adding a potential only in a subspace preserves the momentum symmetry partially:
\begin{\eq}
\int_{M_{\rm sub}} dt d^d x  V( \phi ) ,
\end{\eq}
where $M_{\rm sub}$ is a subspace.
For this case, we still have the subsystem momentum symmetries
associated with the charges that do not touch points in $M_{\rm sub}$.
It would be interesting to find a systematic classification
of interactions preserving the subsystem symmetries.

\subsubsection{Lattice theory}
The $\phi$-theory was originally constructed from lattice theories 
called ``XY-plaquette model" in $d=2$ case \cite{Seiberg:2020bhn} or ``XY-cube model" in $d=3$ case 
\cite{Gorantla:2020xap}.
Here we discuss a lattice theory corresponding to the $\phi$-theory in generic dimensions.
Let us consider a $d$-dimensional hypercubic lattice with a spacing $a$ and 
label each site by 
$\vec{n}=(n_1 , \cdots ,n_d )$ $\in$ $ \mathbb{Z}_{N_1} \times \cdots \times \mathbb{Z}_{N_d}$.
Generalizing the $d=2$ and $d=3$ cases,
we define the lattice Hamiltonian as
\begin{\eq}
H = \frac{\mu_0}{2a^d} \sum_{\vec{n}} (\pi_{\vec{n}})^2 
+\frac{\mu_0 \alpha^2 }{a^d} \sum_{\vec{n}}  \left( 1 -\mathrm{cos} (\Delta_\#^d \phi_{\vec{n}})  \right) , 
\end{\eq}
where $\Delta_\#$ is a difference operator defined as
\begin{\eq}
\Delta_\#^d := \prod_{i=1}^d \Delta_i, \quad 
\Delta_i \phi_{\vec{n}} := \phi_{\vec{n}+\vec{e}_i} - \phi_{\vec{n}} .
\end{\eq}
The canonical commutation relation is
\begin{\eq}
[\phi_{\vec{n}},\pi_{\vec{n}'} ] = i \delta_{\vec{n},\vec{n}'}.
\end{\eq}
One can easily show that
if we set
\begin{\eq}
\phi_{\vec{n}} \rightarrow \phi(x), \quad \frac{\pi_{\vec{n}}}{a^d} \rightarrow \Pi(x) ,
\end{\eq}
and take the $a\rightarrow 0$ limit,
then the Hamiltonian becomes the one \eqref{eq:Hamiltonian_phi} of the continuous $\phi$-theory.

The lattice model also has subsystem symmetries which become the momentum multi-pole symmetries \eqref{eq:trans_boson}
in the continuum limit, as in the $d=2$ and $d=3$ cases.
Those symmetries are associated with the following conserved charges
\begin{\eq}
Q^i_{(\vec{n} ;\hat{n}^i )}
:= \sum_{n^i} \pi_{\vec{n}} ,
\end{\eq}
where the symbol $(\vec{n} ;\hat{n}^i )$ denotes dependence on the site $\vec{n}$ except the $i$-th direction.
This operator commutes with the Hamiltonian and generates the following $U(1)$ rotation of $\phi_{\vec{n}}$:
\begin{\eq}
Q^i_{(\vec{n} ;\hat{n}^i )} : 
\phi_{\vec{n}} \ \rightarrow\ \phi_{\vec{n}}  +f^i_{(\vec{n} ;\hat{n}^i )} ,
\end{\eq}
where $f^i_{(\vec{n} ;\hat{n}^i )} \in [0,2\pi )$ independent of $n^i$
satisfying $\Delta_i f^i_{(\vec{n} ;\hat{n}^i )} =0$.
This is interpreted as a lattice counterpart of the momentum multi-pole symmetries \eqref{eq:trans_boson}.
As in the continuum theory,
all the $Q^i_{(\vec{n} ;\hat{n}^i )}$'s are not independent of each other and
we have the constraint
\begin{\eq}
\sum_{n^j}  Q^i_{(\vec{n} ;\hat{n}^i )} = \sum_{n^i}  Q^j_{(\vec{n} ;\hat{n}^j )} ,
\end{\eq}
which is parallel to \eqref{eq:Gauss}.

\subsection{Chiral $\phi$-theory}
\label{sec:chiral-phi}
As we have seen,
the $\phi$-theory is a generalization of the standard compact boson in $1+1$ dimensions and
has some similar properties. 
Recently it was also found that
there is a chiral boson-like counterpart of the $\phi$-theory for $d=3$, 
which is called ``chiral $\phi$-theory" \cite{Burnell:2021reh,Yamaguchi:2021xeq}
although a precise meaning of ``chiral" is currently mysterious.
Here we consider such a theory for general odd $d$.

We define the Lagangian of the chiral $\phi$-theory in $d+1$ dimensions as 
\begin{\eq}
\mathcal{L}_\pm 
:= \frac{\mu_c}{2} (\del_\mp \phi ) (\delS \phi ) ,
\end{\eq}
where the spatial dimension $d$ is odd\footnote{
For even $d$, we naively have $\int dt d^d x (\del_t \phi) (\delS \phi ) =0$ and
it does not seem to be interpreted as an analogue of chiral boson.
}. 
The constant $\mu_c$ can be arbitrarily real at this stage while we will take it to be properly quantized in the anomaly inflow argument discussed in sec.~\ref{sec:chiral_inflow}.
The equation of motion of this theory is given by
\begin{\eq}
\del_\mp \delS \phi =0 .
\label{eq:EOMchi}
\end{\eq}
As for the $\phi$-theory, the equation of motion \eqref{eq:EOMchi} 
can be regarded as a conservation law of subsystem symmetries:
\begin{\eq}
\del_t P_t = \delS P_\# ,
\end{\eq}
where
\begin{\eq}
P_t := \mu_c \delS \phi, \quad P_\# := \pm \mu_c \alpha \delS \phi.
\end{\eq}
The corresponding charges are different between $\alpha \neq 0$ and $\alpha =0$ for $d\neq 1$.
For $\alpha \neq 0$, it is
\begin{\eq}
Q^i_m(x;\hat{x}^i) := \int dx^i P_t ,
\label{eq:momentum}
\end{\eq}
which is similar to the ordinary $\phi$-theory.
For $\alpha =0$, 
the conserved charge is the local operator $P_t (x) $ itself
since the equation of motion is simply $\del_t P_t =0$.
We will discuss this point in more detail in sec.~\ref{sec:alpha0}.

The theory also has the winding symmetry coming from the trivial identity
\begin{\eq}
\del_t W^\#_t - \del_\#^d W = 0 ,
\end{\eq}
where
\begin{\eq}
W^\#_t := \frac{1}{2\pi} \del_\#^d \phi, \quad W := \frac{1}{2\pi} \del_t \phi.
\end{\eq}
The corresponding subsystem charges are
\begin{\eq}
Q^i_w(x;\hat{x}^i) := \int dx^i W^\#_t.
\end{\eq}
Note that this is proportional to the charge $Q^i_m(x;\hat{x}^i)$ of the momentum symmetry in \eqref{eq:momentum}
since the time components of both the currents are proportional to each other.
The other components are different
but they become essentially the same after using the equation of motion.
Therefore the two symmetries seem to be equivalent at least classically in the chiral $\phi$-theory
while 
the momentum symmetry current
may change upon adding interactions.

One can see that 
the action is also invariant under the following transformation
\begin{\eq}
\phi(t,x)\ \rightarrow\ \phi(t,x) + f^i (t,x;\hat{x}^i) ,
\end{\eq}
which is interpreted as a gauge symmetry.
This can be also seen from the fact that
any modes satisfying $\delS \phi =0$ (i.e.~momentum modes) do not contribute to the action.
Therefore the momentum modes can be gauged away by the above transformation
and unphysical in the chiral $\phi$-theory in contrast to the non-chiral $\phi$-theory.

\subsubsection{Vacuum structure on torus}
Let us switch to the operator formalism and study the vacuum structures on torus.
The conjugate momentum of $\phi$ is given by
\begin{\eq}
\Pi_\pm := \frac{\del \mathcal{L}_\pm}{\del (\del_t \phi )}
=\frac{\mu_c}{2} \del_\#^d \phi .
\end{\eq}
Then the Hamiltonian of the chiral $\phi$-theory is
\begin{\eq}
\mathcal{H}_\pm 
= \pm \frac{2\alpha}{\mu_c} \Pi_\pm^2 ,
\end{\eq}
with the canonical commutation relation 
\begin{\eq}
[\phi (x) , \Pi_\pm (y) ] = i \delta^{(d)} (x-y ) .
\end{\eq}
Since the conjugate momentum is a space derivative of $\phi$,
this also implies
\begin{\eq}
[\phi (x) ,\phi (y) ] = i (-1)^d \frac{2}{\mu_c} \theta^{(d)} (x-y)  ,
\end{\eq}
which reflects a non-local nature of the chiral $\phi$-theory
similar to the usual chiral boson.
Since the momentum modes are unphysical,
the vacuum of the chiral $\phi$-theory on torus is unique.
In the infinite volume limit,
the modes with non-zero $p_1 \cdots p_d$ can have energies smoothly connected to the ground state energy and
hence the theory is gapless.

\subsubsection{'t Hooft anomaly and anomaly inflow}
\label{sec:chiral_inflow}
As in the chiral boson, we can see that the chiral $\phi$-theory has a 't Hooft anomaly,
It was also shown in \cite{Burnell:2021reh} that
the chiral $\phi$-theory for $d=3$ and $\alpha =0$ has a similar 't Hooft anomaly.
Here we discuss this point for general odd $d$. 
For this purpose, let us consider the Euclidean action:
\begin{\eq}
\mathcal{L}_\pm = \frac{\mu_c}{2} \left( i\del_\tau \phi \mp \alpha \delS \phi \right) (\delS\phi) .
\end{\eq}
The current of the momentum subsystem symmetry is
\begin{\eq}
P_\tau = i \mu_c \delS \phi, \quad P_\# = \mp \mu_c \alpha \delS \phi ,
\end{\eq}
with the conservation law
\begin{\eq}
\del_\tau P_\tau + \delS P_\# = 0.
\end{\eq}

To study possible 't Hooft anomalies,
let us promote the momentum symmetry transformation 
to a shift of a function with a full spacetime dependence:
\begin{\eq}
\phi \rightarrow \phi + f(\tau,x).
\end{\eq}
We can couple the current $P_\tau$ to a U(1) background tensor gauge field $(A_\tau, A_\#)$,
which transforms as
\begin{\eq}
A_\tau \rightarrow A_\tau + \del_\tau f(\tau,x), \quad A_\# \rightarrow A_\# + \del_\#^d f(\tau,x).
\end{\eq}
The Lagrangian with a minimal coupling to the background gauge field is
\begin{\eq}
\mathcal{L}_\pm [A]
= \frac{\mu_c}{2}\left(i\del_\tau \phi \del_\#^d \phi \mp \alpha \del_\#^d \phi \del_\#^d \phi \right) - A_\tau P_\tau - A_\# P_\# +\frac{\mu_c}{2} \left(i A_\tau A_\# \mp \alpha A_\# A_\# \right) ,
\end{\eq}
where the last two terms are counter terms and they do not affect 't Hooft anomalies.
The variation of the gauged action is 
\begin{\eq}
\delta S_\pm 
= -i\frac{\mu_c}{2} \int d\tau d^{d}x f(\del_\tau A_\# - \del_\#^d A_\tau) ,
\label{eq:anomaly_chiral}
\end{\eq}
which signals a 't Hooft anomaly.

The 't Hooft anomaly can be compensated by a ``Chern-Simons theory" in one higher dimensions as shown in \cite{Burnell:2021reh} for the $\alpha =0$ and $d=3$ case.
Here we consider generic $\alpha$ and odd $d$. 
The Lagrangian of the ``Chern-Simons theory" is
\begin{\eq}
\mathcal{L}_{\mathrm{CS}}^c 
= \frac{i\mu_c}{2}\left[ A_\tau (\del_z A_\# - \delS A_z)
- A_z (\del_\tau A_\# - \delS A_\tau)
+ A_\# (\del_\tau A_z - \del_z A_\tau) \right] ,
\end{\eq}
where gauge invariance requires $\mu_c$ to be quantized as $\mu_c =\mathbb{Z}/2\pi$.
If we put the theory on a space with a boundary at $z=0$,
the action under the gauge transformation changes as
\begin{\eq}
\delta S_{\rm CS}^c
= \frac{i\mu_c}{2} \int_{z=0} d\tau d^d x f(\del_\tau A_\# - \del_\#^d A_\tau) ,
\end{\eq}
which compensates the 't Hooft anomaly \eqref{eq:anomaly_chiral}.

One may wonder if the winding symmetry also has a 't Hooft anomaly but there is a subtlety as follows.
Let us first gauge the winding symmetry
by introducing a background gauge field $(\tilde{A}_\tau, \tilde{A}_\#)$:
\begin{\eq}
\phi \rightarrow \phi + \tilde{f}(\tau,x), \quad \tilde{A}_\tau \rightarrow \tilde{A}_\tau + \del_\tau \tilde{f}(\tau,x), \quad \tilde{A}_\# \rightarrow \tilde{A}_\# + \del_\#^d \tilde{f}(\tau,x).
\end{\eq}
Then the Lagrangian with a minimal coupling to the background gauge field is
\begin{\eq}
\mathcal{L}_\pm [\tilde{A}]
= \frac{\mu_c}{2}\left( 
i\del_\tau \phi \del_\#^d \phi \mp \alpha \del_\#^d \phi \del_\#^d \phi - 2\pi \tilde{A}_\tau W^\#_\tau - 2\pi \tilde{A}_\# W \right) ,
\end{\eq}
where
\begin{\eq}
W_\tau^\# = \frac{i}{2\pi} \delS \phi, \quad W = -\frac{i}{2\pi}\del_\tau \phi .
\end{\eq}
The variation of the action under the background gauge transformation is
\begin{\eq}
\delta S 
= \frac{\mu_c}{2} \int dt d^d x 
\left( 2i \del_\tau \phi \delS \tilde{f} + i \del_\tau \tilde{f} \delS \tilde{f} \mp 2\alpha \delS \tilde{f} \delS \phi 
\mp \alpha \delS \tilde{f} \delS \tilde{f} - i \tilde{A}_\tau \delS \tilde{f} + i \tilde{A}_\# \del_\tau \tilde{f} \right) ,
\end{\eq}
which is nonzero.
One might attempt to interpret it as a 't Hooft anomaly
in the sense that the action becomes non-gauge invariant once we turn on the background gauge field. 
However it seems difficult to interpret as a usual 't Hooft anomaly
since the variation depends not only on the background gauge field but also on $\phi$
while the quadratic term in $\tilde{f}$ can be cancelled by adding appropriate counter terms.
This subtlety appears also in the standard chiral boson but we have not found any literature mentioning this. It would be nice if one can find some interpretation.

\subsubsection{Comments on symmetries in $\alpha =0$}
\label{sec:alpha0}
As we have mentioned,
the corresponding charges associated with the momentum symmetry
are different between $\alpha \neq 0$ and $\alpha =0$.
Let us see this point in more detail.

For $\alpha \neq 0$, the conserved charges are
\[
Q^i(x;\hat{x}^i) = \int dx^i P_t.
\]
Taking a linear combination of these charges as
\begin{\eq}
\int d^{d-1}x f^i(x; \hat{x}^i) Q^i(x;\hat{x}^i) ,
\end{\eq}
we see that this generates the following transformation
\begin{\eq}
\phi \rightarrow \phi + \sum_{i=1}^d f^i (x;\hat{x}^i).
\end{\eq}
In contrast,
the conservation law for $\alpha =0$ is
\begin{\eq}
\del_t P_t = 0 \quad  {\rm for}\ \alpha =0 ,
\end{\eq}
which indicates that the local operator $P_t$ itself is conserved. 
Then, the linear combination
\begin{\eq}
\int d^d x f(x) P_t ,
\end{\eq}
generates the symmetry transformation
\begin{\eq}
\phi \rightarrow \phi + f(x) ,
\end{\eq}
which is larger than the $\alpha\neq 0$ case.
This situation is similar to symmetries in quantum mechanics and $d$-form symmetries in $d+1$ dimensional field theories.
Note that the difference here is in contrast to the $d=1$ case,
where the parameter $\alpha$ can be absorbed by a coordinate transformation\footnote{
There is a possibility that a similar coordinate transformation exists also for $d\neq 1$
while we have not found.
}.

\section{Supersymmetric and fermionic theories with subsystem symmetries}
\label{sec:SUSY}
In this section,
we construct supersymmetric and fermionic theories with subsystem symmetries in $d+1$ dimensions with odd $d$
by extending the construction of \cite{Yamaguchi:2021qrx} in $3+1$ dimensions.

\subsection{Construction of supersymmetric theory}
We construct a SUSY theory with subsystem symmetries 
whose SUSY structure is similar to $\mathcal{N}=(1,1)$ SUSY theory in $1+1$ dimensions.
First let us introduce the following ``superspace" 
spanned by
\begin{\eq}
( t, x_1 , \cdots , x_d ) \quad {\rm and}  \quad (\theta^+ , \theta^- ) ,
\end{\eq}
where $\theta^\pm$ is a one-component real fermionic coordinate.
In terms of the superspace coordinate, we define the differential operators
\begin{\eq}
\mathcal{Q}_\pm :=  -i \frac{\del}{\del \theta^\pm} +2\theta^\pm \del_\pm ,\quad
\mathcal{D}_\pm :=  -i \frac{\del}{\del \theta^\pm} -2\theta^\pm \del_\pm ,
\end{\eq}
which satisfy
\begin{\eqa}
&& \{ \mathcal{Q}_\pm , \mathcal{Q}_\pm \} =-4i\del_\pm ,\quad 
\{ \mathcal{Q}_+ ,\mathcal{Q}_- \} =0 ,\NN\\
&& \{ \mathcal{D}_\pm ,\mathcal{D}_\pm \} = 4i\del_\pm ,\quad 
\{ \mathcal{D}_+ ,\mathcal{D}_- \} =0 ,\quad
\{ \mathcal{Q}_\alpha , \mathcal{D}_\beta \} =0 .
\end{\eqa}
We also introduce a real superfield:
\begin{\eq}
\Phi (t, x , \theta^+ ,\theta^- ) 
:= \phi (t, x ) +i\theta^+ \psi_+ (t,x) +i\theta^- \psi_- (t,x) +i\theta^+ \theta^- f(t,x) .
\end{\eq}
We define SUSY transformation for superfield as
\begin{\eq}
\delta \Phi =\mathbb{Q}\Phi
:= (i\epsilon_- \mathcal{Q}_+ -i\epsilon_+ \mathcal{Q}_- ) \Phi .
\end{\eq}
For each component of the superfield, we have the SUSY transformation
\begin{\eqa}
\delta \phi &=& i\epsilon_- \psi_+ -i\epsilon_+ \psi_- ,\NN\\
\delta \psi_+ &=& -2\epsilon_- \del_+ \phi -\epsilon_+ f ,\NN\\
\delta \psi_- &=& 2\epsilon_+ \del_- \phi -\epsilon_- f ,\NN\\
\delta f &=& 2i\epsilon_- \del_+ \psi_- +2i\epsilon_+ \del_- \psi_+ .
\end{\eqa}

To construct a SUSY invariant action,
we note that
given two superfields $\Phi_1 ,\Phi_2 $, 
the combination $(\delta \Phi_1 )\Phi_2 + \Phi_1 (\delta \Phi_2 )$ is total derivative in superspace for odd $d$.
The difference between odd $d$ and even $d$ essentially comes 
from the ``Leibniz rule" \eqref{eq:Leibniz} for $\delS$.
Using this fact, we find
\begin{\eqa}
&& \delta \int dt d^d x \int d^2 \theta \Phi_1 \Phi_2  \NN\\
&=& \int dt d^d x \int d^2 \theta \Bigl[ 
(\delta \Phi_1 )\Phi_2 + \Phi_1 (\delta \Phi_2 ) \Bigr] \NN\\
&=& 2i \int dt d^d x d^2 \theta 
\Bigl[ \epsilon_-  \theta^+ \left\{ 
  (\del_+ \Phi_1 )\Phi_2 + \Phi_1 (\del_+  \Phi_2 ) \right\}
-\epsilon_+  \theta^-  \left\{ 
  (\del_- \Phi_1 )\Phi_2 + \Phi_1 (\del_-  \Phi_2 ) \right\} \Bigr]  \NN\\
&=& 0 .
\end{\eqa}
This leads us to a systematic way to construct a SUSY invariant action.
In particular let us consider the following Lagrangian
\begin{\eq}
\mathcal{L}_{\rm SUSY}
= \frac{\mu_0}{2} \int d^2 \theta  \mathcal{D}_- \Phi \mathcal{D}_+ \Phi .
\end{\eq}
Noting 
\begin{\eqa}
\mathcal{D}_\pm \Phi
=  \psi_\pm -2\theta^\pm \del_\pm \phi  \pm \theta^\mp f
 \mp 2i\theta^+ \theta^- \del_\pm \psi_\mp ,
\end{\eqa}
the Lagrangian in the component representation is
\begin{\eq}
\mathcal{L}_{\rm SUSY}
=  \frac{\mu_0}{2} \Bigl[  4 (\del_- \phi ) (\del_+ \phi) +2i \psi_+ \del_- \psi_+ +2i \psi_- \del_+ \psi_- +f^2  \Bigr] .
\label{eq:L_SUSY}
\end{\eq}
The first term is nothing but the Lagrangian of the $\phi$-theory.
The second and third terms look kinetic terms of the fermions but with the exotic derivative $\delS$.
This can be regarded as a fermionic version of the $\phi$-theory and called ``$\psi$-theory" \cite{Yamaguchi:2021qrx}.
Thus the supersymmetric theory consists of the $\phi$-theory and $\psi$-theory while they are decoupled.

\subsection{Subsystem symmetries}
\label{sec:SUSY_sym}
The supersymmetric theory also has subsystem symmetries.
First we obviously have the subsystem symmetries of the $\phi$-theory:
the multi-pole momentum and winding symmetries.
In particular here it is convenient to write the conservation law for the momentum symmetry as
\begin{\eq}
\del_\mp P_\pm =0 , \quad P_\pm := \mu_0 \del_\pm \phi  .
\end{\eq}
Let us focus on symmetries coming from the $\psi$-theory part.
The equation of motion for $\psi_\pm$ is
\begin{\eq}
\del_\mp \psi_\pm =0 .
\label{eq:eom_fermion}
\end{\eq}
This is interpreted as the conservation law for the current of ``fermionic multi-pole momentum symmetry",
all of whose components are $\psi_\pm$.
An associated conserved charge with this is
\begin{\eq}
Q^i_\pm (x;\hat{x}^i )
= \mu_0 \oint dx^i \psi_\pm ,
\label{eq:charge_fermion}
\end{\eq}
which generates a shift of $\psi_\pm$:
\begin{\eq}
Q_\pm^i (x;\hat{x}^i ) : \psi_\pm (t,x)\ \rightarrow\ 
\psi_\pm (t,x) +\chi_\pm^i (x;\hat{x}^i ).
\label{eq:trans_fermi}
\end{\eq}

It is interesting to note that
the current for the fermionic momentum symmetry is a super-partner of the bosonic momentum symmetry \cite{Yamaguchi:2021qrx}. 
Indeed we can write the conservation laws in a unified way:
\begin{\eq}
\mathcal{D}_\mp \mathcal{J}_\pm =0,
\end{\eq}
where
\begin{\eq}
\mathcal{J}_\pm 
:= \mu_0 \mathcal{D}_\pm \Phi
=  \mu_0 (\psi_\pm -2\theta^\pm \del_\pm \phi  \pm \theta^\mp f
 \mp 2i\theta^+ \theta^- \del_\pm \psi_\mp ).
\label{eq:current_SUSY}
\end{\eq}
So $\mathcal{J}_\pm $ can be regarded as a current multiplet for the momentum symmetries.

\subsection{Self-duality}
As for the $\phi$-theory,
we can show that the SUSY theory has a self-duality
by an argument parallel to sec.~\ref{sec:duality_phi}.
We start with the original Lagrangian in superspace:
\begin{\eq}
\mathcal{L}_{\rm SUSY}^{(0)} [\Phi ]
= \frac{\mu_0}{2}  \mathcal{D}_- \Phi \mathcal{D}_+ \Phi .
\end{\eq}
where the dynamical variable is the real superfield $\Phi$.
Next we switch the dynamical variable to $(\mathcal{D}_+ \Phi ,\mathcal{D}_- \Phi)$ rather than $\Phi$
by introducing a Lagrange multiplier to impose the constraint
\begin{\eq}
 \mathcal{D}_+ (\mathcal{D}_- \Phi )  +\mathcal{D}_- (\mathcal{D}_+ \Phi ) =0 ,
\end{\eq}
which is an extension of the conservation law of the winding symmetry.
This is realized by the Lagrangian 
\begin{\eq}
\mathcal{L}_{\rm SUSY}^{(1)} [ \mathcal{D}_+ \Phi , \mathcal{D}_- \Phi ,\tilde{\Phi} ]
=  \frac{\mu_0}{2}  \mathcal{D}_- \Phi \mathcal{D}_+ \Phi  
-c\tilde{\Phi} \left\{  \mathcal{D}_+ (\mathcal{D}_- \Phi )  +\mathcal{D}_- (\mathcal{D}_+ \Phi ) \right\} ,
\end{\eq}
where $\tilde{\Phi} \sim \tilde{\Phi} +2\pi $ is a periodic real superfield  
playing the role of the Lagrange multiplier.
The coefficient $c$ is determined by imposing invariance of the action 
under the transformation $\tilde{\Phi} \rightarrow \tilde{\Phi} +2\pi $ as
\begin{\eq}
c= \frac{1}{4\pi \alpha} .
\end{\eq}
Up to total derivative terms,
we can rewrite the Lagrangian as
\begin{\eq}
\mathcal{L}_{\rm SUSY}^{(1)}  [ \mathcal{D}_+ \Phi , \mathcal{D}_- \Phi ,\tilde{\Phi} ]
\simeq  \frac{\mu_0}{2}  \mathcal{D}_- \Phi \mathcal{D}_+ \Phi  
+\frac{1}{4\pi \alpha}  \left\{  ( \mathcal{D}_+ \tilde{\Phi} )  (\mathcal{D}_- \Phi )  
 +( \mathcal{D}_- \tilde{\Phi} )  (\mathcal{D}_+ \Phi ) \right\} .
\end{\eq}
Then integrating $\mathcal{D}_\pm \Phi$ out leads us to
\begin{\eq}
\mathcal{D}_\pm \Phi 
= \pm \frac{1}{2\pi \alpha \mu_0} \mathcal{D}_\pm \tilde{\Phi} ,
\label{eq:SUSY_map}
\end{\eq}
and we finally obtain 
\begin{\eqa}
\mathcal{L}_{\rm SUSY}^{(2)} [ \tilde{\Phi} ]
= \frac{\tilde{\mu}_0}{2} ( \mathcal{D}_- \tilde{\Phi} )  (\mathcal{D}_+ \tilde{\Phi} ) , 
\end{\eqa}
where
\begin{\eq}
\tilde{\mu}_0 = \frac{1}{(2\pi \alpha)^2 \mu_0 } .
\end{\eq}
This is nothing but the original Lagrangian with a replacement $\mu_0 \rightarrow \tilde{\mu}_0$.
Thus the SUSY theory also has the self-duality.

Let us also see a correspondence between the symmetries. 
The equation of motion for $\tilde{\Phi}$ is
\begin{\eq}
\mathcal{D}_- \mathcal{D}_+ \tilde{\Phi} =0 ,
\end{\eq}
which is interpreted as the conservation law for the momentum symmetry in the dual theory.
The duality map \eqref{eq:SUSY_map} tells us that
this equation holds in the original theory regardless of the equation of motion:
\begin{\eq}
\mathcal{D}_- \mathcal{D}_+ \tilde{\Phi} 
=\frac{1}{2} \left\{ \mathcal{D}_- ( \mathcal{D}_+ \tilde{\Phi} )
  -\mathcal{D}_+ (\mathcal{D}_-   \tilde{\Phi} ) \right\} 
=\frac{1}{4\pi \alpha \mu_0} \left\{ \mathcal{D}_- ( \mathcal{D}_+ \Phi )
  +\mathcal{D}_+ (\mathcal{D}_- \Phi ) \right\}
=0 ,  
\end{\eq}
which implies that this is the conservation law of the winding symmetry in the original theory.
By a similar argument,
we can easily show that
the conservation law for the winding symmetry in the dual theory
is equivalent to the equation of motion in the original theory. 
Therefore, the duality map exchanges the momentum and winding symmetries
as in the $\phi$-theory.

\subsection{Gauging momentum symmetries with supersymmetry}
Let us gauge the momentum symmetries preserving the supersymmetry.
For this purpose, it is convenient to work in the superfield formalism.
From \eqref{eq:trans_boson} and \eqref{eq:trans_fermi},
we see that the bosonic and fermionic momentum symmetries transform the superfield $\Phi$ as
\begin{\eq}
Q_m^i (x;\hat{x}^i ), Q_\pm^i (x;\hat{x}^i ) :
\Phi  \rightarrow 
\Phi + \Bigl[ f^i (x;\hat{x}^i ) \ 
+i\theta^+ \chi_+^i (x;\hat{x}^i )  +i\theta^- \chi_-^i (x;\hat{x}^i )  +i\theta^+ \theta^- \times 0 \Bigr] ,
\end{\eq}
which is a shift by a superfield independent of $x^i$.
To gauge the symmetries, we promote it to a shift by a generic real superfield:
\begin{\eq}
\Phi \quad \rightarrow\quad \Phi + K ,
\label{eq:promote}
\end{\eq}
where $K$ is a real bosonic superfield with the periodicity $K \sim K+2\pi$.
Then the derivative $\mathcal{D}_\pm \Phi$ is no longer invariant under the shift \eqref{eq:promote} and
we introduce a new dynamical degrees of freedom to compensate the variance as in usual gauge theory.
Specifically we introduce a fermionic real superfield $\Gamma_\pm$ 
transforming under the shift as
\begin{\eq}
\Gamma_\pm \quad \rightarrow \quad \Gamma_\pm +\mathcal{D}_\pm K ,
\end{\eq}
which is the gauge transformation in the superfield language.
Using this, we define a supercovariant derivative $\nabla_\pm$ as
\begin{\eq}
\nabla_\pm \Phi 
:= \mathcal{D}_\pm \Phi -\Gamma_\pm ,
\end{\eq}
which is invariant under the shift \eqref{eq:promote}.
Then introducing
\begin{\eq}
\Sigma = \frac{i}{2} (\mathcal{D}_+ \Gamma_- +\mathcal{D}_- \Gamma_+ ) ,
\end{\eq}
the kinetic term for the gauge field is given by
\begin{\eq}
\mathcal{L}_{\rm kin}
:=\frac{1}{g^2} \int d^2 \theta\ \mathcal{D}_- \Sigma \mathcal{D}_+ \Sigma ,
\end{\eq}
which is supersymmetric by construction.
To write down this in components, we expand the superfield $\Gamma_\pm$ as
\begin{\eq}
\Gamma_\pm 
= \chi_\pm -2\theta^\pm A_\pm \pm\theta^\mp (B \pm \sigma ) 
 \mp 2i\theta^+ \theta^- (\lambda_\pm +\del_\pm \chi_\mp ) ,
\end{\eq}
where $B$, $\sigma$ and $A_\pm$ are real bosonic fields 
while $\chi_\pm$ and $\lambda_\pm$ are real femionic fields.
Then taking an analogue of Wess-Zumino gauge as
\begin{\eq}
\chi_\pm =0 ,\quad B=0 ,
\end{\eq}
we find
\begin{\eq}
\mathcal{L}_{\rm kin}
=\frac{4}{g^2} \Biggl[ F_{+-}^2 +(\del_+ \sigma )(\del_- \sigma ) 
+\frac{i}{2} \lambda_+ \del_- \lambda_+ +\frac{i}{2} \lambda_- \del_+ \lambda_- \Biggr] ,
\end{\eq}
where 
\begin{\eq}
F_{+-} := \del_+ A_- -\del_- A_+ .
\end{\eq}
We can also add other SUSY invariant terms for the gauge field.
For example, we have an analogue of a theta-term in $(1+1)$-dimensional $U(1)$ gauge theory:
\begin{\eq}
\mathcal{L}_{\rm theta}
:= \int d^2 \theta \left( -\frac{i\theta}{2\pi \alpha} \Sigma \right)
= -\frac{\theta}{\pi\alpha} F_{+-} .
\end{\eq}
The kinetic term for the matters can be made gauge invariant by using the supercovariant derivative as
\begin{\eq}
\frac{\mu_0}{2} \int d^2 \theta\ \nabla_- \Phi \nabla_+ \Phi .
\end{\eq}

\subsection{Purely fermionic theory}
\label{sec:fermion}
Let us pick up the fermion part of the above SUSY theory:
\begin{\eq}
\mathcal{L}_\psi
:= \frac{\mu_0}{2}\Bigl[ i \psi_+ (\del_t - \alpha \delS ) \psi_+
 + i \psi_- (\del_t + \alpha \delS ) \psi_- \Bigr] .
\label{eq:L_psi}
\end{\eq}
This is a purely fermionic theory and 
a generalization of the ``$\psi$ theory'' introduced in \cite{Yamaguchi:2021qrx} for $d=3$ to general odd $d$.
As we discussed in sec.~\ref{sec:SUSY_sym},
the equation of motion \eqref{eq:eom_fermion} for the fermions
can be interpreted as the conservation law of the subsystem symmetries 
with the charges \eqref{eq:charge_fermion}.
For $d=1$, the theory describes the two free chiral fermions with $\pm$ chiralities.
For $d>1$, the differential operators in the Lagrangian look similar to the ones of the chiral fermions 
but it is unclear 
whether there is a generalization of the concept of chirality in this context.
It would be interesting if one can explore this aspect in more detail.
In addition, we could also add some interactions among fermions and bosons as in \cite{Distler:2021bop}.
It would be also nice to classify possible interactions (partially) preserving subsystem symmetries.

\subsubsection{Boundary conditions}
If we put the theory on torus $S_{L_1}^1 \times \cdots \times S_{L_d}^1$,
then we have to specify a boundary condition as in usual fermions.
A significant feature in this type of theory is that
such a boundary condition is generically position-dependent:
\begin{\eq}
\left. \psi_\pm (t, x) \right|_{x^i \rightarrow x^i +L_i } 
=\sigma_\pm^i (x;\hat{x}^i ) \psi (t, x)  
\quad \left( \sigma_\pm^i (x;\hat{x}^i ) = \pm 1 \right) ,
\end{\eq}
because the theory allows discontinuous configurations of the fields
whose discontinuity is independent of one of the directions.
Furthermore, 
since the sign $\sigma_\pm^i (x;\hat{x}^i )$ is position-dependent,
we have to also specify a boundary condition for that:
\begin{\eq}
\left. \sigma_\pm^i (x;\hat{x}^i ) \right|_{x^j \rightarrow x^j +L_j } 
= \sigma_\pm^{ij} (x;\hat{x}^i ,\hat{x}^j ) \sigma_\pm^i (x;\hat{x}^i )  
\quad \left( \sigma_\pm^{ij} (x;\hat{x}^i ,\hat{x}^j ) = \pm 1 \right) ,
\end{\eq}
as noted in \cite{Cao:2022lig} for $(2+1)$-dimensional theories.
Note that the sign $\sigma_\pm^{ij} (x;\hat{x}^i ,\hat{x}^j )$ is
symmetric in the indices $(i,j)$ 
since
\begin{\eqa}
&& \left. \psi_\pm (t, x)  
\right|_{x^i \rightarrow x^i +L_i ,x^i \rightarrow x^j +L_j }  \NN\\
&=& \left. \sigma_\pm^i (x;\hat{x}^i ) \psi (t, x)  \right|_{\rightarrow x^j +L_j }
= \sigma_\pm^{ij} (x;\hat{x}^i ,\hat{x}^j ) 
  \sigma_\pm^i (x;\hat{x}^i ) \sigma_\pm^j (x;\hat{x}^j ) \psi (t, x)  \NN\\
&=& \left. \sigma_\pm^j (x;\hat{x}^j ) \psi (t, x)  \right|_{\rightarrow x^i +L_i }
= \sigma_\pm^{ji} (x;\hat{x}^i ,\hat{x}^j ) 
  \sigma_\pm^i (x;\hat{x}^i )  \sigma_\pm^j (x;\hat{x}^j ) \psi (t, x) . \NN
\end{\eqa}
This kind of procedures are repeated until we specify boundary conditions along all the circles.
Note that 
the boundary condition affects amount of the subsystem symmetries \eqref{eq:trans_fermi} generated by $Q_\pm^i (x;\hat{x}^i)$;
if we take an anti-periodic boundary condition at $x$ 
along the $i$-th circle $S^1_{L_i}$,
then we no longer have the subsystem symmetry for those $i$ and $(x;\hat{x}^i)$
since the anti-periodic boundary condition does not allow
Fourier zero modes along $S^1_{L_i}$ at that point.
In addition, for the case of the SUSY theory,
if the set of boundary conditions includes one anti-periodic boundary condition at least,
then we also loose the supersymmetry.
These are parallel to usual fermions subject to anti-periodic boundary conditions.

\subsubsection{Comments on lattice regularization}
We can also consider a lattice regularization of the $\psi$-theory.
As ordinary fermionic field theories have subtleties in regularizing massless fermions,
the $\psi$-theory also has similar subtleties
as pointed out in \cite{Yamaguchi:2021qrx} for the $d=3$ case.

To see this, let us consider an analogue of ``naive" fermion in ordinary lattice field theories. 
We put a real fermion field $c_{\vec{n}}$ on sites of the lattice that
satisfies the canonical anti-commutation relation
\begin{\eq}
\{ c_{\vec{n}} ,c_{\vec{n}'} \} = \delta_{\vec{n},\vec{n}'} .
\end{\eq}
We consider the following Hamiltonian  
\begin{\eq}
H_{\rm naive}
=  i\beta \sum_{\vec{n}} c_{\vec{n}} \tilde{\Delta}_\#^d c_{\vec{n}} ,
\end{\eq}
where $\beta := \pm \mu_0 \alpha /a^d$ and 
\begin{\eq}
\tilde{\Delta}_\#^d := \prod_{i=1}^d \tilde{\Delta}_\#^i ,\quad
\tilde{\Delta}_\#^i f_{\vec{n}} := f_{\vec{n}+e_i} -f_{\vec{n}-e_i} .
\end{\eq}
To see the vacuum structure of the theory,
let us consider the Fourier expansion
\begin{\eq}
c_{\vec{n}} 
= \frac{1}{\sqrt{V}} \sum_{\vec{k}} b_{\vec{k}} e^{i\vec{k}\cdot (a\vec{n})} ,
\end{\eq}
where $V:=\prod_{i=1}^d L_i$ and
\begin{\eq}
\{ b_{\vec{k}} ,b_{\vec{k}'} \} = \delta_{\vec{k}, -\vec{k}'} .
\end{\eq}
Then the Hamiltonian becomes
\begin{\eq}
H_{\rm naive}
= \beta \sum_{\vec{k}} \left( \prod_{i=1}^d \sin{(k_i a)} \right) b_{-\vec{k}} b_{\vec{k}}.
\end{\eq}
Therefore the condition for zero modes is 
\begin{\eq}
\prod_{i=1}^d \sin{(k_i a)}  = 0 ,
\label{eq:lattice_zero}
\end{\eq}
in contrast to the ordinary naive lattice fermion
with the condition $\sum_{i=1}^d \sin{(k_i a)}  = 0$.
The condition \eqref{eq:lattice_zero} is satisfied when one of $k_i$'s at least is either 0 or $\pi /a$ rather than all the $k_i$'s.
Therefore the number of the zero modes is
\begin{\eq}
2 \sum_{i=1}^d  \frac{V}{L_i}
-4 \sum_{1\leq i<j \leq d}  \frac{V}{L_i L_j}
+\cdots 
+2^{d-2} \sum_{1\leq i<j \leq d}  L_i L_j
-2^{d-1} \sum_{i=1}^d L_i
+2^d ,
\end{\eq}
while the correct number approaching the original continuum theory should be
\begin{\eq}
 \sum_{i=1}^d  \frac{V}{L_i}
- \sum_{1\leq i<j \leq d}  \frac{V}{L_i L_j}
+\cdots 
+ \sum_{1\leq i<j \leq d}  L_i L_j
- \sum_{i=1}^d L_i
+1 ,
\label{eq:modes_correct}
\end{\eq}
since it should come from only the modes such that one of $k_i$'s at least is 0.
Thus we have ``doublers" in much more complicated way than ordinary lattice fermions.

In principle there may be various ways to circumvent the ``doubling problem"
for the $\psi$-theory.
Here we consider an analogue of the Wilson fermion prescription:
\begin{\eq}
H_{\rm Wilson}
=  i\beta \sum_{\vec{n}} \Biggl[ c_{\vec{n}} \tilde{\Delta}_\#^d c_{\vec{n}} ,
-r  c_{\vec{n}} \Delta_\#^{2d} c_{\vec{n}} \Biggr] ,
\end{\eq}
where $\Delta_\#^{2d}:=(\Delta_\#^d )^2$.
In terms of the Fourier modes, it is written as 
\begin{\eq}
H_{\rm Wilson}
= \beta \sum_{\vec{k}} \left( \prod_{i=1}^d \sin{(k_i a)}
-r \prod_{i=1}^d \left( 1-\cos{(k_i a)} \right) \right) b_{-\vec{k}} b_{\vec{k}}.
\end{\eq}
Then the condition for zero modes is
\begin{\eq}
\prod_{i=1}^d \sin{(k_i a)} -r \prod_{i=1}^d \left( 1-\cos{(k_i a)} \right) =0 .
\end{\eq}
For $r\neq 0$, the second term lifts
the ``doubler modes" in the naive fermion, 
where one of the $k_i a$'s at least is $\pi$.
Therefore the number of the zero modes is
the correct number \eqref{eq:modes_correct} approaching the original continuum theory.
It would be interesting to look for
other ways to avoid the doubling problem 
such as analogues of Staggered, overlap and domain wall fermions.
It would be also illuminating to see
whether there is an analogue of the Nielsen-Ninomiya theorem \cite{Nielsen:1980rz,Nielsen:1981xu}
for field theories with subsystem symmetries.

\section{Chiral supersymmetric and fermionic theories with subsystem symmetries}
\label{sec:chiralSUSY}
In this section
we supersymmetrize the chiral $\phi$-theory.
As we will see,
this is quite parallel to the supersymmetrization of the usual chiral boson in $1+1$ dimensions with $\mathcal{N}=(1,0)$ SUSY.
To our knowledge, such a generalization to $d\neq 1$ has not been done in literature.
As in the $\psi$-theory,
a purely fermionic theory can be obtained by dropping the bosonic part of the SUSY theory.

\subsection{Supersymmetric ``chiral"  theory}
As we have seen in sec.~\ref{sec:chiral-phi},
the Lagrangian of the chiral $\phi$-theory is
\[
\mathcal{L}_\pm 
= \frac{\mu_c}{2} (\del_\mp \phi ) (\delS \phi ) ,
\]
Let us add a fermion to this theory and consider the Lagrangian
\begin{\eq}
\mathcal{L}_{\rm SUSY}^\pm 
=\frac{\mu_c}{2} \Bigl[ \del_\#^d \phi \del_\mp\phi +\frac{i}{2}  \psi_\pm \del_\mp \psi_\pm \Bigr] .
\label{eq:chiralSUSY}
\end{\eq}
We can easily show that 
the action is invariant under the following fermionic transformation:
\begin{\eqa}
\delta \phi &=& -i\epsilon \psi_\pm , \NN\\
\delta \psi_\pm &=& -2\epsilon \del_\#^d \phi. 
\end{\eqa}
Therefore we interpret this theory as a supersymmetrization of the chiral $\phi$-theory.

One can also describe the above theories in terms of superfields
in a similar way to the superfield formalism for $1+1$ dimensional $\mathcal{N}=(1,0)$ theories \cite{Sonnenschein:1988ug}.
Let us introduce the differential operators 
\begin{\eq}
\mathcal{Q}:= i \frac{\del}{\del \zeta^\pm} -2\zeta^\pm \delS , \quad
\mathcal{D}:= i \frac{\del}{\del \zeta^\pm} +2\zeta^\pm \delS ,
\end{\eq}
which satisfy
\begin{\eq}
\mathcal{Q}\mathcal{Q} =-2i\delS ,\quad \mathcal{D}\mathcal{D} = 2i\delS ,\quad  
\{\mathcal{Q} ,\mathcal{D} \} =0 .
\end{\eq}
We also introduce a real superfield $\Phi_\pm$ as
\begin{\eq}
\Phi_\pm = \phi + i\zeta^\pm \psi_\pm ,
\end{\eq}
and define a SUSY transformation as
\begin{\eq}
\delta \Phi_\pm = i \epsilon \mathcal{Q}\Phi .
\end{\eq}
Noting
\begin{\eq}
\mathcal{D} \Phi_\pm = -\psi_\pm +2\zeta_\pm \delS \phi ,
\end{\eq}
we can rewrite the supersymmetric Lagrangian \eqref{eq:chiralSUSY} as
\begin{\eq}
\mathcal{L}_{\rm SUSY}^\pm 
= \frac{\mu_c}{4}\int d^2 x d\zeta^\pm \del_\mp \Phi_\pm \mathcal{D}\Phi_\pm .
\end{\eq}
The equation of motion in the language of the superfield is
\begin{\eq}
\del_\mp \mathcal{D} \Phi_\pm = 0 .
\end{\eq}
This is interpreted as the following conservation law
\begin{\eq}
\del_\mp \mathcal{J}_c^\pm = 0 ,
\end{\eq}
where 
\begin{\eq}
\mathcal{J}_c^\pm := \mu_c \mathcal{D}\Phi_\pm .
\end{\eq}
The top component of $\mathcal{J}_c^\pm$ is 
the same as the charge \eqref{eq:momentum} of the momentum symmetry in the chiral $\phi$-theory
while the bottom component is the charge associated with a shift of $\psi_\pm$.
Therefore $\mathcal{J}_c^\pm$ is interpreted as a supersymmetric current multiplet of the momentum symmetries
similar to \eqref{eq:current_SUSY}.

\subsection{Purely fermionic ``chiral"  theory}
As in sec.~\ref{sec:fermion},
dropping the chiral $\phi$-theory part of the chiral SUSY theory leads us to 
the following purely fermionic theory:
\begin{\eq}
\mathcal{L}_\psi^\pm 
=\frac{i\mu_c}{4}   \psi_\pm \del_\mp \psi_\pm  .
\end{\eq}
For $d=1$, this is nothing but the single free chiral fermion with $\pm$ chirality.
This theory can be also obtained by dropping either $\psi_+$ or $\psi_-$ in the $\psi$-theory \eqref{eq:L_psi}.
We call this theory ``chiral $\psi$-theory".
As in the $\psi$-theory, when we put the theory on a torus,
there are infinitely many choices of boundary conditions for the fermion.
Similarly, taking a periodic boundary condition along all the circles at every point
preserves the full shift symmetry
while the symmetry is reduced for anti-periodic boundary conditions.
The SUSY in the chiral SUSY theory is also lost
if we take an anti-periodic boundary condition along one direction at one point at least.

\section{Vacuum structures of theories with fermions}
\label{sec:vacuum}
In this section, we discuss vacuum structures of the theories with fermions
introduced in the previous sections.
We will see that an essence of our argument is nicely captured 
by the free massless Dirac fermion.
Therefore we will start with a problem to determine the vacuum structure of the free massless Dirac fermion as a warm-up 
and then generalize it to the theories with the subsysmtem symmetries.

\subsection{Warm-up: free massless Dirac fermion}
\label{sec:warm-up}
Let us consider a $(d+1)$-dimensional free massless Dirac fermion on torus $T^d$.
The Lagrangian and Hamiltonian are given by
\begin{\eq}
\mathcal{L}_{\rm Dirac}
=i\bar{\psi}\gamma^\mu \del_\mu \psi ,
\end{\eq}
and
\begin{\eq}
H_{\rm Dirac}
=\int d^d x  \left( -i\bar{\psi}\gamma^j \del_j \psi \right) ,
\end{\eq}
respectively.
Degeneracy of the vacua in this theory is of course well known:
when the fermion is subject to the periodic boundary condition along all the spatial directions,
it is $2^{[\frac{d+1}{2}]}$ the same as the number of components of the Dirac spinor
while otherwise it is unique.
The difference essentially comes from the fact that
an anti-periodic boundary condition prohibits the fermion to have zero modes.
In the infinite volume limit,
the non-zero modes have energies smoothly connected to the ground state energy and
the theory is gapless.

Let us discuss the above structure from the viewpoint of symmetries.
When the ferimion satisfies the periodic boundary condition along all the spatial directions,
the theory has fermionic shift symmetry:
\begin{\eq}
\psi \longrightarrow \psi + \xi ,\quad
\bar{\psi} \longrightarrow \bar{\psi} + \tilde{\xi} ,
\label{eq:shift_Dirac}
\end{\eq}
where $\xi$ and $\tilde{\xi}$ are constant grassmann spinors.
These symmetries gives the fermionic Noether currents:
\begin{\eq}
J^\mu = -i\bar{\psi} \gamma^\mu , \quad 
\tilde{J}^\mu = i \gamma^0 \gamma^\mu \psi ,
\end{\eq}
whose conservation laws are equivalent to the Dirac equation.
Their corresponding Noether charges are
\begin{\eq}
Q = \int d^d x J^0 = -i \int d^d x \psi^\dagger , \quad
\tilde{Q} = \int d^d x \bar{J^0} = i \int d^d x \psi .
\end{\eq}
We can see that these two charges do not commute each other:
\begin{\eq}
\left\{ Q_\alpha , \tilde{Q}_\beta \right\} = \int d^dx \int d^dy \{\psi_\alpha^\dag (x),\psi_\beta (y)\} 
= i \delta_{\alpha\beta}\cdot {\rm vol}(T^d ) ,
\end{\eq}
where $\alpha$ and $\beta$ run $1,\cdots , 2^{[\frac{d+1}{2}]}$.

The above commutation relation implies that the vacuum structure cannot be trivial: 
if we assume that a ground state $|\Omega\rangle$ is invariant under $Q_\alpha$
i.e.~$Q_\alpha |\Omega \rangle$ =0,
the commutation relation shows that
$\tilde{Q}_\alpha |\Omega \rangle$ cannot be proportional to $|\Omega \rangle$.
In addition, the state $\tilde{Q}_\alpha |\Omega \rangle$ is also a ground state
since $\tilde{Q}_\alpha$ commutes with the Hamiltonian.
A similar thing holds if we interchange $Q_\alpha$ and $\tilde{Q}_\alpha$ in the above argument.
Hence the ground states of this theory must satisfy either
$Q_\alpha \left| \Omega \right> \neq 0$ or $\tilde{Q}_\alpha \left| \Omega\right> \neq 0$ for every $\alpha$.
This implies that 
a half of the fermionic shift symmetries \eqref{eq:shift_Dirac} at least are
spontaneously broken and
the theory has degenerate vacua.
Thus degeneracy of the vacua is $2^{[\frac{d+1}{2}]}$ at least.

\subsection{$\psi$-theory}
Let us consider the $\psi$-theory 
on torus $S^1_{L_1}\times \cdots \times S^1_{L_d}$:
\[
\mathcal{L}_\psi
= i \psi_+ \del_- \psi_+ + i \psi_- \del_+ \psi_- .
\]
As noted in sec.~\ref{sec:fermion},
we have infinitely many choices for boundary conditions of the fermions
but let us first take a periodic boundary condition along all the directions
at all the points.
The Hamiltonian of this theory is
\begin{\eq}
H_\psi
= \int d^d x \Bigl[ i \psi_+ \delS \psi_+  - i \psi_- \delS \psi_-  \Bigr] ,
\end{\eq}
where the fermion field operators satisfy the commutation relations
\begin{\eq}
\{\psi_\pm(t,x), \psi_\pm(t,y)\} = \delta^{(d)} (x-y), 
\quad \{\psi_\pm(t,x), \psi_\mp(t,y)\} =0 .
\label{eq:canonical_psi}
\end{\eq}
The vacuum structures turn out to be clearer if we make a Fourier expansion
\begin{\eq}
\psi_\pm (t,x)
=\frac{1}{(2\pi)^d} \sum_{k_1 ,\cdots ,k_d \in \mathbb{Z}} \psi_{\pm ,k} (t) e^{ik\cdot x}  .
\end{\eq}
We can easily see that the zero modes of $\delS$ with $k_1 \cdots k_d =0 $ do not contribute to the Hamiltonian.
This implies that the vacua of the $\psi$-theory are degenerate
since acting the zero mode operators on a ground state generate other ground states.

We can see the above structures from the viewpoint of the symmetries as in the last subsection.
As argued in sec.~\ref{sec:SUSY_sym},
the theory has the subsystem symmetry charges
\[
Q^i_\pm(x;\hat{x}^i) = \oint dx^i \psi_\pm(x) ,
\]
which generates the fermionic shift:
\[
\psi_\pm(x) \longrightarrow \psi_\pm(x) + \chi_\pm (x;\hat{x}^i) .
\]
While this is more complicated than the shift symmetry \eqref{eq:shift_Dirac} 
for the massless Dirac fermion, 
we can do a similar argument to study vacuum structures 
based on commutation relations among the charges.
Using the commutation relations \eqref{eq:canonical_psi}, we find
\begin{\eq}
\begin{split}
\{Q^i_\pm(x;\hat{x}^i), Q^j_\pm(y;\hat{y}^j)\}
&=\oint dx^i \oint dy^j \{\psi_\pm(x), \psi_\pm(y)\} \\
&=\oint dx^i \oint dy^j  \delta^{(d)} (x-y) \\
&=\left\{ 
\begin{aligned}
& L_i \delta^{(d-1)}(x-y; \hat{i}) \quad \mathrm{for} \; i=j \\
& \delta^{(d-2)}(x-y; \hat{i}, \hat{j}) \quad \mathrm{for} \; i \neq j
\end{aligned}
\right. ,
\end{split} 
\end{\eq}
and
\begin{\eq}
\{Q^i_\pm(x;\hat{x}^i), Q^j_\mp(y;\hat{y}^j)\} = 0 ,
\end{\eq}
where $\delta^{(d-1)}(x; \hat{i}) $ denotes
the product of the delta functions of the directions 
except the $i$-th direction.
In particular, the above relations tell us that
square of the charge $Q^i_\pm (x;\hat{x}^i)$ cannot be zero for any $x$ and $i$.
Therefore upon acting on a ground state $|\Omega \rangle$,
we must satisfy
\begin{\eq}
 Q^i_\pm (x;\hat{x}^i) |\Omega \rangle \neq 0 ,\quad
{\rm for}\  ^\forall x\ {\rm and}\ i .
\end{\eq}
This indicates that the subsystem symmetries are spontaneously broken and 
the theory has infinitely many vacua
since $Q^i_\pm (x;\hat{x}^i) |\Omega \rangle$ also has the same energy as the ground state.
Therefore the vacuum Hilbert space should form a non-trivial representation of the symmetries.
Roughly speaking,
the degeneracy of vacua is proportional to the number of points of $d-1$ dimensional torus
although we need a regularization to be more precise.
So far we have taken the boundary condition to be periodic along all the directions
at all the points.
If we take an anti-periodic boundary condition along one of the circles at one point,
then it reduces the degeneracy by one.

\subsection{Non-chiral supersymmetric theory}
Next let us consider the non-chiral SUSY theory constructed in sec.~\ref{sec:SUSY},
whose Lagrangian is
\[
\mathcal{L}_{\rm SUSY} 
=\frac{\mu_0}{2} \Bigl[ 4 (\del_- \phi ) (\del_+ \phi) 
+2i \psi_+ \del_- \psi_+ +2i \psi_- \del_+ \psi_- +f^2 \Bigr] . 
\]
Since all the fields are decoupled,
the vacua of this theory are simply tensor products of
the vacua of the $\phi$-theory and $\psi$-theory.
Therefore, the degeneracy of the vacua in the SUSY theory is simply 
the same as the one of the $\psi$-theory and 
the fermionic momentum symmetries are spontaneously broken.

An extra aspect in the SUSY case is of course the presence of the supercharges:
\begin{\eq}
\mathcal{Q}_\pm
= \int d^d x \psi_\pm \del_\pm \phi .
\end{\eq}
Then it would be natural to ask whether the SUSY is preserved on the vacua.
To see this, we compute an algebra 
among the supercharges and the charges of the fermionic momentum symmetries:
\begin{\eq}
\{ \mathcal{Q}_\pm  , Q_\pm^i (x;\hat{x}^i ) \}
= \mu_0 \int d^d y \int dx^i \{ \psi_\pm (y) \del_\pm \phi (y) , \psi_\pm (x) \}
= \mu_0 \int dx^i  \del_\pm \phi (x) 
= Q_m^i (x;\hat{x}^i ) .
\end{\eq}
Since the vacua have the vanishing charge of the bosonic momentum symmetry,
this implies that a ground state $|\Omega \rangle$ satisfies
\begin{\eq}
\mathcal{Q}_\pm  \left(  Q_\pm^i (x;\hat{x}^i ) |\Omega \rangle \right) 
= \frac{1}{2}  Q_m^i (x; \hat{x}^i )  |\Omega \rangle = 0 ,
\end{\eq}
meaning that the vacua preserve the supersymmetry.

\subsection{Chiral $\psi$-theory}
The Lagrangian of the chiral $\psi$-theory is
\[
\mathcal{L}_\psi^\pm 
=\frac{i\mu_c}{4}   \psi_\pm \del_\mp \psi_\pm  .
\]
This theory is only the $\psi_+$ or $\psi_-$ part of the $\psi$-theory.
Therefore we can repeat the argument of the $\psi$-theory,
focusing only on the subsystem symmetry charge $Q_+^i (x;\hat{x}^i )$ or $Q_-^i (x;\hat{x}^i )$.
Thus the fermionic momentum symmetry is spontaneously broken also in the chiral $\psi$-theory.

\subsection{Chiral supersymmetric theory}
The Lagrangion of the chiral supersymmetric theory is 
\[
\mathcal{L}_{\rm SUSY}^\pm 
=\frac{\mu_c}{2} \Bigl[ \del_\#^d \phi \del_\mp\phi +\frac{i}{2}  \psi_\pm \del_\mp \psi_\pm \Bigr] .
\]
As in the non-chiral SUSY theory,
all the fields are decoupled and
hence the vacua of the chiral SUSY theory are simply tensor products of
the ones of the chiral $\phi$-theory and chiral $\psi$-theory.
Therefore, the degeneracy of the vacua is  
the same as the one of the chiral $\psi$-theory.
The fermionic momentum symmetries are spontaneously broken on the vacua and
the vacuum Hilbert space forms a non-trivial representation of those symmetries 
associated with the symmetry breaking.

Let us see a property of the vacua under the supersymmetry generated 
by the supercharge
\begin{\eq}
\mathcal{Q}
= \int d^d x \psi_\pm \delS \phi .
\end{\eq}
A straightforward computation shows that the supercharge satisfies the algebra
\begin{\eqa}
\left\{ \mathcal{Q}  , Q_\pm^i (x;\hat{x}^i ) \right\}
= Q_m^i (x;\hat{x}^i ) .
\end{\eqa}
Then, 
one can show that a ground state $|\Omega \rangle$ satisfies
\begin{\eq}
\mathcal{Q}  \left(  Q_\pm^i (x;\hat{x}^i ) |\Omega \rangle \right) 
= \frac{1}{2}  Q_m^i (x; \hat{x}^i )  |\Omega \rangle = 0 ,
\end{\eq}
as in the non-chiral SUSY theory.
This implies that the vacua of the chiral SUSY theory preserve the supersymmetry.

One might wonder if an algebra argument for the vacuum structures could give
some nontrivial structures for the bosonic momentum symmetries
because of the commutation relation unique for the chiral theory:
\begin{\eq}
[\Pi_\pm (x),\Pi_\pm (y)] = i \partial_\#^d \delta^{(d)} (x-y).
\end{\eq}
This is not the case
because the charges of the bosonic momentum symmetries do not have non-trivial algebra:
\begin{\eq}
\left[ Q^i_m(x;\hat{x}^i), Q^j_m(x;\hat{x}^j) \right]
=\frac{4}{\mu_c^2} \oint dx^i \oint dy^j \left[ \Pi_\pm (x),\Pi_- (y)\right] 
= \frac{4i}{\mu_c^2}\oint dx^i \oint dy^j \partial_\#^d \delta^d(x-y) 
= 0 .
\end{\eq}
This is indeed consistent with the fact that
the bosonic momentum symmetries are not spontaneously broken in the chiral $\phi$-theory. 

\section{Conclusion and discussions}
\label{sec:conclusion}
In this paper 
we studied the various non-relativistic field theories with the subsystem symmetries.
We started with the $\phi$-theory in $d+1$ dimensions
and discussed its properties studied in literature for $d\leq 3$
such as symmetries, self-duality, vacuum structures, 't Hooft anomaly and lattice regularization.
Next we studied the chiral $\phi$-theory
which is a chiral boson like counterpart of the $\phi$-theory.
Then we turned to theories with fermions.
We first constructed the supersymmetric version of the $\phi$-theory,
and dropping the bosonic part of the supersymmetric theory,
found the purely fermionic theory with the subsystem symmetries called $\psi$-theory.
We argued that the naive lattice regularization of the $\psi$-theory
suffers from an analogue of the doubling problem as pointed out in the $d=3$ case \cite{Yamaguchi:2021qrx}.
To avoid the problem,
we proposed an analogue of Wilson fermion which makes the ``doubling modes" infinitely massive in the continuum limit.
We also constructed the supersymmetric chiral $\phi$-theory.
We finally discussed vacuum structures of the theories including fermions and 
found that the vacua are infinitely degenerate.

There are various interesting directions for future.
We have seen that
the models discussed in this paper have common features to ordinary $(1+1)$-dimensional field theories.
Therefore it would be interesting to see
whether or not there are other common phenomena
such as bosonization, anomalies and index theorem.
We also discussed the analogues of the chiral boson and chiral fermion
while a precise meaning of ``chirality" in this context is still unclear.
It would be interesting if one finds an appropriate definition of the ``chirality".

It should be also important to study the ``doubling problem" in lattice regularization of fermionic theories with subsystem symmetries in more detail.
In particular this paper proposed only one prescription by an analogue of Wilson fermion
while there should be other ways to avoid the problem 
such as analogues of Staggered, overlap and domain wall fermions.
It would be also illuminating to see
whether there is an analogue of the Nielsen-Ninomiya theorem \cite{Nielsen:1980rz,Nielsen:1981xu}
for field theories with subsystem symmetries.

Many of ordinary field theories can be realized as worldvolume theories
of D-branes in string theory.
Recently some field theories with subsystem symmetries were constructed 
from brane constructions \cite{Geng:2021cmq}.
It would be interesting to ask whether field theories discussed in this paper or their extensions have connections to string theory
as well as quiver gauge theories \cite{Razamat:2021jkx,Franco:2022ziy} and gravity \cite{Pretko:2017fbf,Benedetti:2021lxj,Benedetti:2022zbb,Hinterbichler:2022agn}.

In some spin models with fracton excitations,
phases are characterized by the foliation structure of spatial manifold \cite{Shirley:2017suz,10.21468/SciPostPhys.6.1.015}.
There have been many works on the field theories which 
describe these phases, which are called ``foliated fracton phases''  \cite{Slagle:2018swq,Slagle:2020ugk,Hsin:2021mjn}.
In recent work \cite{Ohmori:2022rzz}, it was reported that 
one of such field theories have correspondence with the gauge field coming from subsystem symmetries.
It would be important to study about excitations in foliated theories and 
find a correspondence with the theories studied in this paper.
Last but not least,
there should be various interesting directions in future related to other recent progress \cite{Gorantla:2020jpy,Rudelius:2020kta,Gorantla:2021svj,Casalbuoni:2021fel,Angus:2021jvm,Jain:2021ibh,Gorantla:2021bda,Gorantla:2022eem,Jensen:2022iww,Perez:2022kax,Hirono:2022dci,Gorantla:2022mrp,2022arXiv220907987E,Gorantla:2022ssr,Gorantla:2022pii}.

\subsection*{Acknowledgement}
We thank Kenya Ikeda and Satoshi Yamaguchi for useful discussions.
M.~H. is supported by MEXT Q-LEAP and JST PRESTO Grant Number JPMJPR2117, Japan.
M.~H. and T.~N. are supported by JSPS Grant-in-Aid 
for Transformative Research Areas (A) ``Extreme Universe" JP21H05190 [D01].

\bibliographystyle{utphys}
\bibliography{fracton}

\end{document}